\documentclass{elsart1p}

\usepackage{graphicx}
\usepackage{url}

\begin{document}

\begin{frontmatter}

% Title, authors and addresses

% use the thanksref command within \title, \author or \address for footnotes;
% use the corauthref command within \author for corresponding author footnotes;
% use the ead command for the email address,
% and the form \ead[url] for the home page:
% \title{Title\thanksref{label1}}
% \thanks[label1]{}
% \author{Name\corauthref{cor1}\thanksref{label2}}
% \ead{email address}
% \ead[url]{home page}
% \thanks[label2]{}
% \corauth[cor1]{}
% \address{Address\thanksref{label3}}
% \thanks[label3]{}

\title{Pulsed extraction of ionization from helium buffer gas}

% use optional labels to link authors explicitly to addresses:
% \author[label1,label2]{}
% \address[label1]{}
% \address[label2]{}

\author[label1,label2]{D.J.~Morrissey\corauthref{cor1}},
\corauth[cor1]{Corresponding author} \ead{morrissey@nscl.msu.edu}
\author[label1,label3]{G.~Bollen},
\author[label1]{M.~Facina},
\author[label1]{S.~Schwarz}

\address[label1]{National Superconducting Cyclotron Laboratory,}
\address[label2]{Department of Chemistry,}
\address[label3]{Department of Physics and Astronomy,
\\ Michigan State University, East Lansing MI 48824}

\begin{abstract}
% Text of abstract
The migration of intense ionization created in helium buffer gas under the influence of applied electric fields is considered. First the chemical evolution of the ionization created by fast heavy-ion beams is described. Straight forward estimates of the lifetimes for charge exchange indicate a clear suppression of charge exchange during ion migration in low pressure helium.  Then self-consistent calculations of the migration of the ions in the electric field of a gas-filled cell at the National Superconducting Cyclotron Laboratory (NSCL) using a Particle-In-Cell computer code are presented. The results of the calculations are compared to measurements of the extracted ion current caused by beam pulses injected into the NSCL gas cell.
\end{abstract}

\end{frontmatter}

% main text
\section{Introduction}
\label{sec:intro}

The thermalization of nuclear reaction products in a buffer gas
for subsequent extraction and study is an active area of research.
Themalization provides very-low energy beams of short-lived isotopes
that can be used for precision mass measurements in traps
\cite{MB_PRL08,TE_PRL08,JAC_PRC07},
for laser-spectroscopy\cite{TN_PRA06}, or be (re)accelerated
to higher energies for reaction and nuclear spectroscopic
studies\cite{XW_PAC07,SS_RSI08}.
The products may be from high-energy nuclear reactions with
mass selection before entering the gas \cite{LW_NIM05,GB_PRL06} or be
from low-energy nuclear reactions and fly directly into the
gas \cite{JH_NIM04,GS_NIM03,AT_RSI05,JBN_NIM05,MF_NIM05}. The key
requirement on all of these systems is that the fast reaction
products must be rapidly and efficiently converted into
high-quality, low-energy beams (kinetic energy $\sim$eV). However,
the simple stopping of each nuclear recoil in the buffer gas
creates from 10$^{4}$ to 10$^{7}$ ion-electron pairs (the primary
ionization) that will hinder the collection of the single incident
ion.

There are a number of detailed reports of the extraction of
nuclear reaction products from gas cells with a variety of
geometries and gas pressures using electric fields
\cite{LW_NIM05,GS_NIM03,AT_RSI05,JBN_NIM05} and without
electrodes\cite{MF_NIM05,JH_NIM04}. The gas cells with electric fields rely
on separating the ion-electron pairs and drifting the cations to the exit
nozzle (the more mobile electrons are expected to be collected rapidly at the
other end of the chamber). These gas cells work in a weak plasma
regime similar to that of an ionization chamber. On the other hand,
the so-called ion guide systems are much smaller and only use gas
flow to extract the ions\cite{JH_NIM04,MF_NIM05}.

\begin{figure}[h] %% 1
\begin{center}
\includegraphics[width=0.45\textwidth]{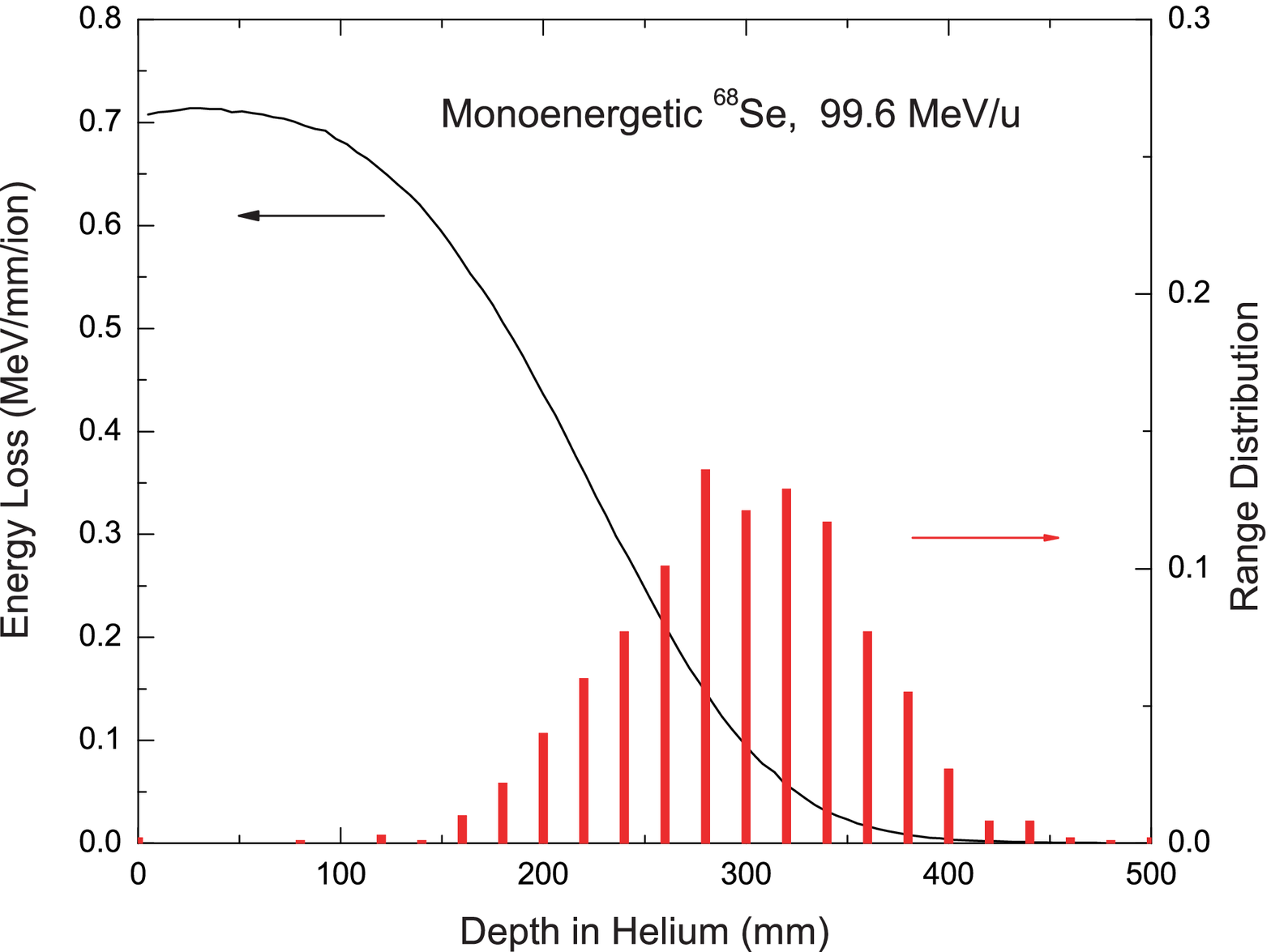}
\includegraphics[width=0.45\textwidth]{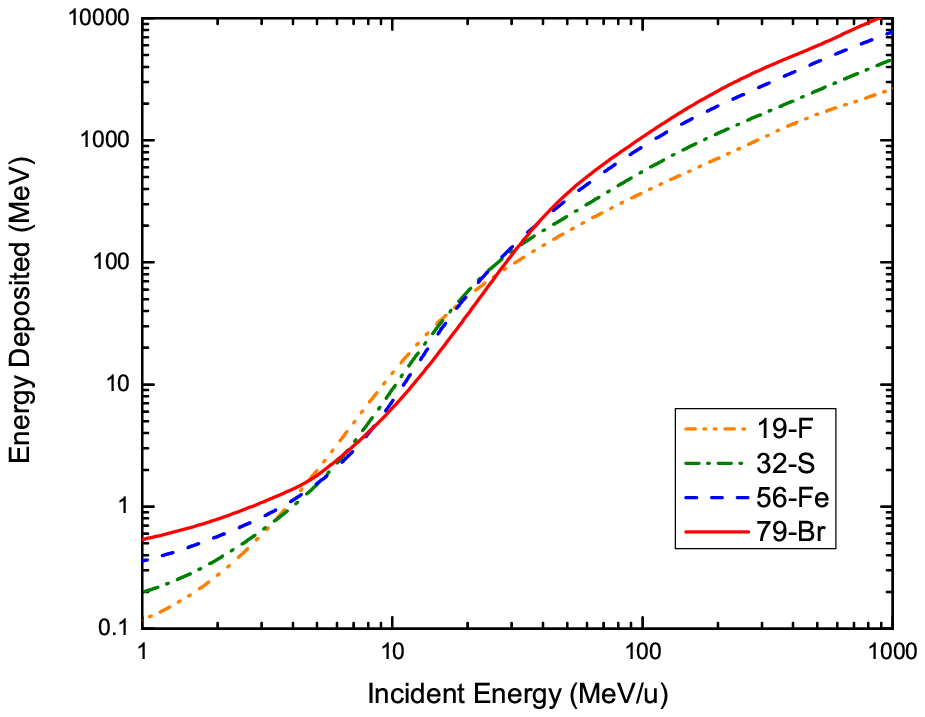}
\caption{\label{fig:sew} (left) The energy loss and range distributions of
monoenergetic $^{68}$Se ions at 99.6~MeV/u that pass through 2.200~mm glass,
0.566~mm beryllium and enter helium gas with a density of 1.65x10$^{19}$~atoms/cm$^3$
calculated with the TRIM program\protect{\cite{TRIM}}. (right)The calculated
residual energy of a fast ion that would be deposited in the helium stopping
gas at the end of the range in a thickness equal to the full-width at
half-maximum of the the range straggling distribution is shown as a
function of the incident kinetic energy.}
\end{center}
\end{figure}

Projectile fragmentation reactions occur at high incident
energies, e.g., $\ge 100$MeV/u and produce a very large variety of
nuclei without decay or chemical losses. The products from these
reactions are used at high energy to carry out much of current
nuclear research \cite{DJM_LNP04}. The stopping and collection of
projectile fragments is highly desirable but particularly
difficult due to their long ranges and concomitant large range
straggling even when range-compression techniques are used
\cite{HW_NIM00,CS_NIM04,LW_NIM04}. For example, the energy loss and
range distributions of energetic $^{68}$Se projectile fragments used in
this work are shown on the left side of Figure~\ref{fig:sew}.  Large range
straggling implies
that the stopped projectile fragments will always have to be collected from
a large spatial region with significant primary ionization.  The calculated
residual energy of a fast ion that would be deposited in 1~bar of helium at
the end of the range in a thickness equal to the full-width at half-maximum
(FWHM) of the the range straggling distributionis shown on the right side
of Figure~\ref{fig:sew}. This estimate was made by obtaining the FWHM of the
range distribution for a monoenergetic ion in helium gas from the well
known \emph{Stopping and Range Tables} \cite{TRIM} and then finding the
kinetic energy of that ion with a range equal to the straggling FWHM.
Figure~\ref{fig:sew} shows
that the increase of deposited energy with incident energy is much
larger than the variation with mass number.  The figure also indicates
that the effects of primary ionization will be much stronger for projectile
fragments produced at 100 MeV/u or more than for nuclear recoils at a
few MeV/u.
%%
%  figure 1 here

Given that the length of a gas cell for efficient collection and the amount
of energy deposited per incident ion are dictated by the incident energy, it
is important to consider the effects of space charge and ion recombination
on the collection of thermalized ions in gas. Huyse
et al. identified the density of ionization created by the stopping ion as an
important parameter to describe the general behavior of the system as
opposed to the simple rate of impinging particles \cite{MH_NIM02}. The
ionization-rate density used in the following discussion is the quotient of
the rate at which ion-electron pairs (initially assumed to be e$^-$ and He$^+$)
are created by the incoming radioactive ions and
the stopping volume.  Recombination processes place an overall
limit on charge buildup in a gas cell. The remaining space charge
affects the ion motion from the point where they stop to the extraction
point as addressed in more detail by Takamine\cite{AT_RSI05}
and Schwarz\cite{SCS_pc05}.  Recombination is not important in
linear gas cells under typical operating conditions and so the
electrons can be rapidly removed leaving
the space-charge from the cations to induce a radial field gradient
that generally pushes the cations out of the ionization volume towards the
walls during their longitudinal drift.

In the present work we describe the measurement and detailed
simulations of cation collection from the NSCL gas cell. The
results from the measurement of the extracted charge with a pulsed
secondary beam were found to be in excellent agreement with
particle-in-cell calculations of the ion migration. After describing the
experiment we address the effects of the chemical evolution of the
initial ionization in helium, an often neglected subject.

\section{Ion Chemistry}\label{sec:ionchem}

As noted above, the slowing down of the nuclear recoils in gas
creates a vast number of ion-electron pairs and the single incident ion is
swept up in the evolution and migration of the buffer gas ions.
The ion currents can be readily measured and used to monitor fast
ion stopping in the gas cell as long as the space charge potential
remains less than the applied potential. In this section, the
general chemical nature of the migrating ions is presented first
and then the level at which ion migration is affected by space
charge along the lines presented by the Leuven
group\cite{MH_NIM02,MF_NIM05} is considered.

Fast nuclear recoils that are thermalized in helium
gas are expected to remain ionized because the first ionization
potential of helium is larger than that of any chemical element.
However, significant chemical evolution of projectile fragments has been
observed with the NSCL gas cell, e.g.,\cite{GB_PRL06,MB_PRL08}
whereas no ion-molecule chemistry and very few stable molecular
ions have been observed to emerge from the SHIPTRAP gas cell
\cite{SAE_NIMB07} even though the state-of-the-art UHV and gas
purification techniques are used in both systems.  The properties
of a prototype cryogenic gas cell were studied by Dendooven and
cryogenic cooling of the gas may be an easier route to ultrahigh
purity gas\cite{PD_NIM06} but no full-scale cryogenic gas cells
exist at present.

While it might be attractive to think that the helium stopping gas
retains the ionization, we will show that the situation is more
complicated.  The initial ion chemistry of helium is somewhat simple
since only three species are possible: He$^+$, its excited states, and
the bare nucleus He$^{2+}$.  As was pointed out some time ago by
Patterson, all of the He$^{2+}$ ions rapidly convert
into He$^+$ by two-body charge transfer reactions and the excited
states of He$^+$ rapidly decay leaving only ground state He$^+$
ions\cite{PLP_PRA70,ECB_PR63}. However, a He$^+$ ion will then
go on to form the relatively stable dimer He$_2^+$ by the reaction:
$$\rm\hspace{2cm} He^+ + He + He \rightarrow He_2^+ + He \hspace{3cm} [1]$$
with a rate given by the expression: $\rm R~=~k_1\rho_n^2$ in
which $\rho_n$ is the number density of (neutral) helium and $\rm
k_1~=~1.08x10^{-31}cm^6/s$ is the second order rate constant.
Details of the formation of the ground and excited states of the
helium molecular ion are not important for the present discussion. It is
easy to show that the half-life for dimer formation is $\rm
T_{1/2}=ln~2/(k_1\rho_n^2)\sim 1\mu$s at a pressure of 100~mbar
and only $\sim$10~ns at 1~bar at ambient temperature.  In fact, it
is very difficult to measure the ion mobility of He$^+$ at
pressures above a few mbar due to their rapid disappearance. Therefore,
all of the primary helium ions produced in gas cells used to
capture nuclear recoils are converted into He$_2^+$ on a time scale
that is small compared to that for ion drift in typical applied
fields (10~V/cm).

\begin{figure}[t] %% 2
\begin{center}
\includegraphics[width=0.95\textwidth]{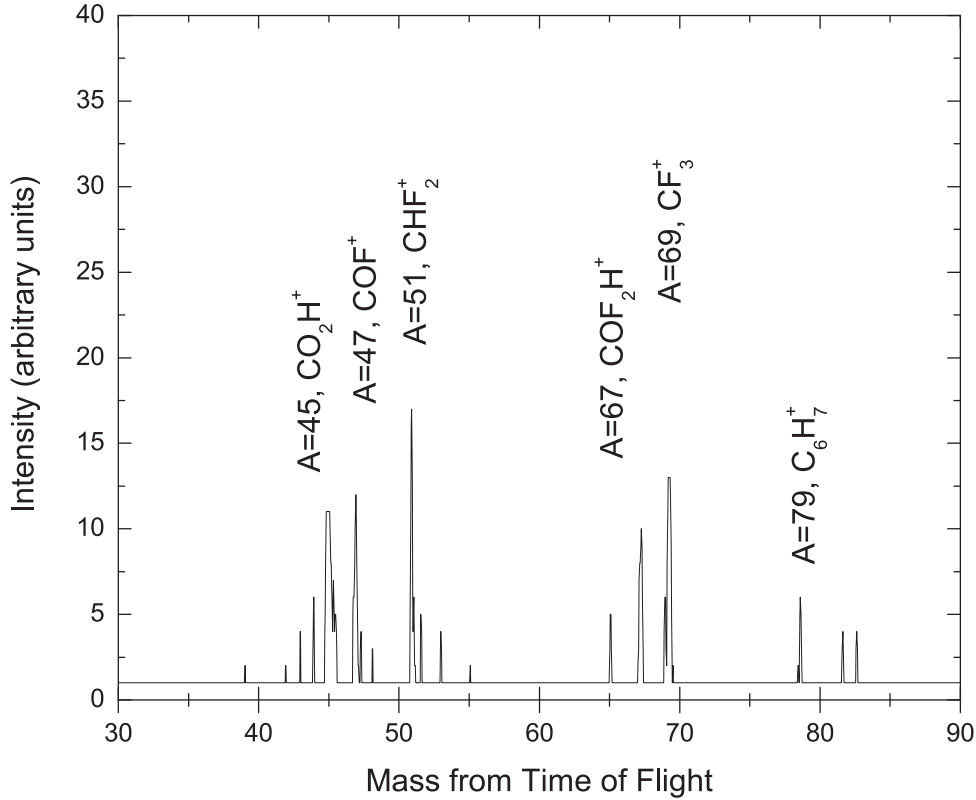}
\caption{\label{fig:mass} The time-of-flight mass spectrum of ions in
the mass range $\sim$40 to $\sim$80 created by an electric discharge
in the NSCL gas cell. Labels indicate specific ions that were identified
by Penning Trap mass spectroscopy.}
\end{center}
\end{figure}

It was also pointed out by Patterson and coworkers that ``small
traces of contaminants [in pure helium gas] become strongly
preferentially ionized by charge transfer.''\cite{ECB_PR63}.
Recent calculations of chemical equilibria in noble gas plasmas
show in detail that trace impurities rapidly become the
significant charge carriers.\cite{TM_APL08} This conversion takes
place by the two-body process (bimolecular reaction):
$$\rm\hspace{2cm} He_2^+ + M \rightarrow He + He + M^+ \hspace{3cm} [2]$$
and is supplemented at high pressures by the three-body process (trimolecular reaction):
$$\rm\hspace{2cm} He_2^+ + M + He \rightarrow He + He + He + M^+ \hspace{3cm} [3]$$
where M is {\em any} neutral impurity molecule.  The chemical nature of
the impurity, M, depends on the history and operation of the gas
cell. For example, extensive studies of the output from the NSCL
gas cell with a Penning trap mass spectrometer have shown that
alkane and alcohol based molecular ions ($\rm C_2H_3\strut^+$,
$\rm C_2H_5\strut^+$, $\rm C_4H_7\strut^+$, $\rm CH_2OH\strut^+$,
$\rm CH_2CH_2OH\strut^+$, etc.) are strongly produced from the gas
cell when it is only cleaned to ``high vacuum'' conditions; that
water-based molecular ions ($\rm H_3O\strut^+$, $\rm
H_5O_2\strut^+$, etc.) are strongly produced shortly after an
ultra-high vacuum (UHV) gas cell is exposed to air; and that
fluorocarbon molecular ions ($\rm CF_3\strut^+$, $\rm
C_2H_3F_2\strut^+$, etc.) are produced by a carefully prepared UHV
gas cell connected to a helium purifier.  For example, the mass spectrum
of ions produced by a discharge in the gas cell just after the beam
time is shown in Figure~\ref{fig:mass}.  Ions with a mass-to-charge ($m/q$)
ratio of approximately 40 to 80 were selected by the ion-guides
and their masses were determined in a time-of-flight measurement.
The prominent ions were identified by observing their resonances
in a Penning Trap mass spectrometer.
%%
%  figure 2 here

\begin{figure}[t] %% 3
\begin{center}
\includegraphics[width=0.95\textwidth]{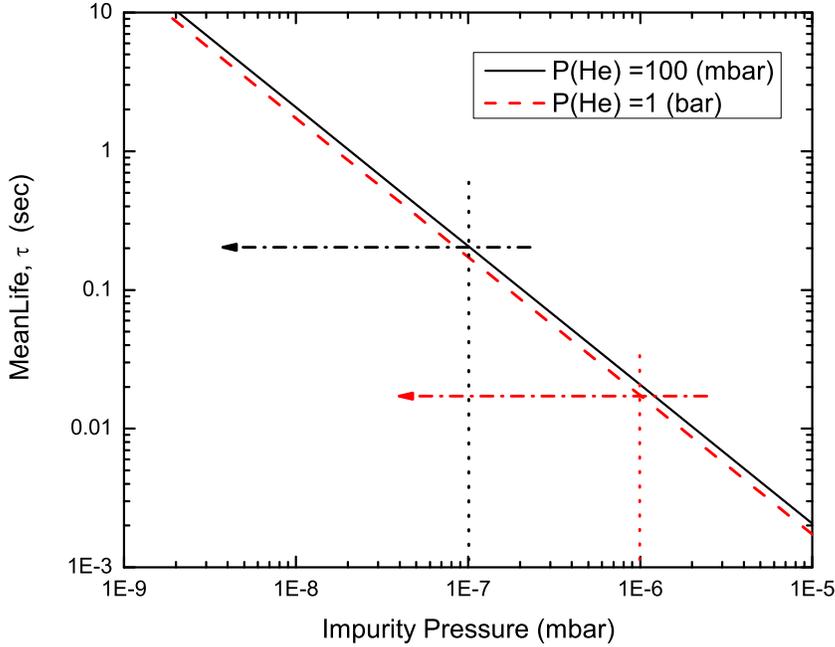}
\caption{\label{fig:tau} The mean lifetime at ambient temperature
for charge exchange between He$_2^+$ and propane (C$_3$H$_8$) as a
function of propane pressure in bulk helium using the rate
constants from Collins\protect{\cite{CBC_JCP78}}.  The dotted
arrows indicate the expected lifetimes for 1~ppb propane in
helium  at two representative pressures.}
\end{center}
\end{figure}

The rate constants for bimolecular charge exchange, k$_2$, (Eq.~2)
are approximately $\rm \sim 2x10^{-9}~cm^3/s$ for a variety of
small molecules including C$_3$H$_8$ and CCl$_2$F$_2$. The rate
constants for trimolecular charge exchange, k$_3$, (Eq.~3) with
these two molecules are approximately $\rm 2x10^{-30}$ and $\rm
8x10^{-30}~cm^6/s$, respectively\cite{CBC_JCP78}.  The mean
lifetime for a He$_2^+$ ion in the presence of trace amounts of
propane (M) is simply $\rm\tau = 1/(k[M])$ where $\rm
k~=~k_2+k_3[He]$ and square brackets indicate concentrations.
The mean lifetime at ambient temperature is shown
in Fig.\ref{fig:tau} as a function of propane pressure in a helium
buffer at pressures of 0.1 and 1~bar. The rates for charge
transfer to other impurities are similar.  Note that charge
transfer takes significantly more time at low impurity levels
than helium dimer formation, thus the charge carriers
evolve from He$^+$ to He$_2^+$ and then to M$^+$ if the ions drift
in the helium for sufficient time.
%%
%  figure 3 here

Currently available helium purifiers and standard UHV techniques
can reduce the impurities in the buffer gas to the level of a few
parts-per-billion (ppb). The mean lifetimes for charge exchange in
helium at this purity level, highlighted in Fig.\ref{fig:tau},
increases from approximately 17~ms to approximately 170~ms as the
pressure decreases from 1.0 to 0.1~bar.  These lifetimes should be
compared to an estimate of the time that nuclear recoils spend
drifting before extraction given by the expression \mbox{$\rm
t~=~\Delta x/(\mu~E)$} where $\Delta$x is the path length, $\mu$ is
the pressure dependent ion mobility, and {\em E} is the net average
electric field. The reason for the difference in the chemical evolution
of the primary ionization in the NSCL and the SHIPTRAP gas cells is
now apparent. The shortest drift time from the midpoint of the NSCL gas
cell at 1~bar of $\rm t\sim 50~ms$ ($\rm\mu\sim20~cm^2/s/V$ and
$E\sim$25~V/cm, see for example, \cite{DAD_NIM06}), is longer than
the mean lifetime for charge exchange at this pressure. On the
other hand, the drift time reported for the SHIPTRAP gas cell
of $\rm t\leq 10~ms$ \cite{SAE_NIMB07} is fifty times shorter than the estimated mean
lifetime for reactions in this system ($\tau\sim$500~ms) at the
same ppb level of contamination.  The very low pressure (40~mbar)
in this device speeds up the drift time and also lowers the rate of
charge exchange of He$_2^+$ with any impurities.

In summary, we expect that all of the primary ionization caused by
the nuclear recoils will be converted into He$_2^+$ ions within a
timescale that is short compared to the extraction time of those
recoils from any gas cell.  The fate of the dimer ions depends on the
level of molecular impurities in the gas and the drift time for
the ions.  Gas purifiers provide a uniform level of impurities
per atom in the ppb range or in other words a relative residual gas pressure
in the helium buffer gas. This uniform purity level does not insure a
uniform amount of charge transfer from the He$_2^+$ ions since
a higher total pressure lowers the ion mobility, increases the
drift time and also increases the charge transfer through an increased
number of ion-molecule collisions.

\section{Experimental}\label{sec:exp}

\begin{figure}[t] %% 4
\begin{center}
\includegraphics[width=0.95\textwidth]{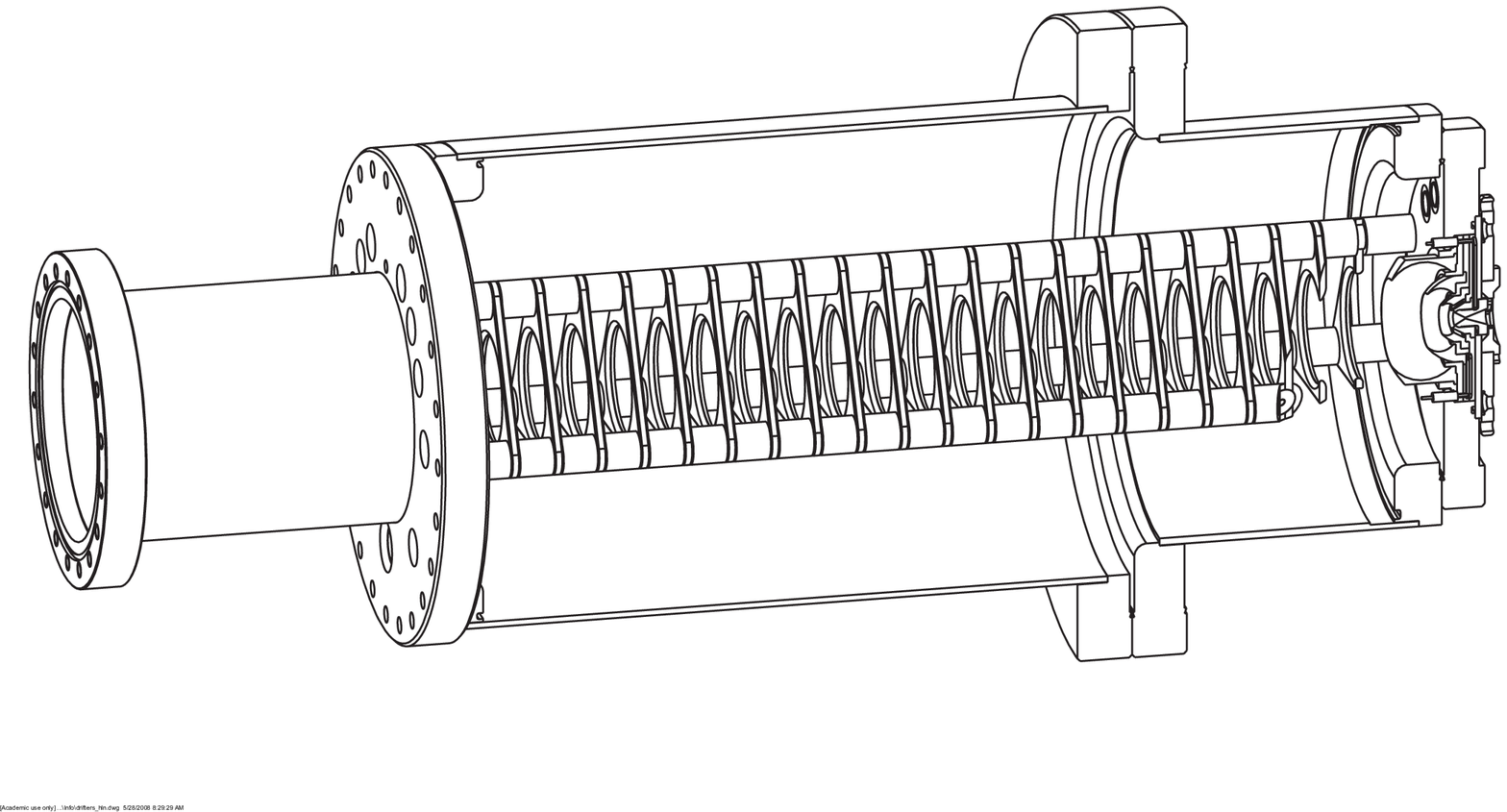}
\caption{\label{fig:gc} A mechanical drawing of the NSCL gas cell with
a cross section showing the drift rings, the spherical focussing electrodes
and the nozzle. Fast projectile fragments enter from the left side
and thermalized ions exit on the right.}
\end{center}
\end{figure}

The present work used the high pressure gas-filled chamber
(so-called ``gas cell'') developed at the NSCL to thermalize,
drift and extract high velocity projectile fragments before they
can decay \cite[and references therein]{LW_NIM05}.  Components of
the fast-ion stopping system include a momentum-dispersive beam
line with precision degraders (1.332~mm SiO$_2$) and a precision
wedge (0.729~mm at midpoint; borosilicate glass) for
momentum compression of the projectile fragments upstream of the
helium-filled gas cell. Degraded but still somewhat energetic ions
enter the gas volume shown in Figure~\ref{fig:gc} (50~cm by 25~cm
diameter) through a polished beryllium window (0.566~mm thick
by 5~cm diameter) and are thermalized in ultra-pure helium gas.
Collinear ring electrodes(inner diameter 5.6~cm) outside the stopping
volume are used to create a DC gradient to drift all cations towards
the exit, at which point four spherical electrodes are used to focus
the ions on the exit nozzle(throat diameter 0.6~mm). Gas flow moves
the ions toward and through the
nozzle into a lower pressure ($\sim 0.1$~mbar) chamber when the
ions approach the throat of the nozzle. A differential pumping system
removes the large amount of neutral helium gas leaving the gas
cell while the ions are captured by a quadrupole ion guide that
extends through the expansion chamber wall into a high vacuum
region ($\sim 10^{-6}$ mbar). The ions are then accelerated up to
5~keV for transport or measurement, see for example
\cite{RR_PRC07,PS_PRC07}.
%%
%  figure 4 here

In the present work, a mixture of isotones with N=34 was
obtained from the A1900 projectile fragment
separator\cite{DJM_NIM03} by the fragmentation of a 151~MeV/u
$^{78}$Kr primary beam in a 331~mg/cm$^2$ beryllium target. The
incident mixed beam contained: $^{65}$Ga, $^{66}$Ge, $^{67}$As,
$^{68}$Se in the proportion of 7:19:10:1 as determined by a
combination of time-of-flight from the production target and
energy loss in a silicon detector that could be positioned inside the
gas cell 39~cm from the entrance. A continuous beam of ions was
delivered by the NSCL facility but short beam pulses (2~ms),
created by manipulating the rf-phase in the K500 accelerator, were also
used to study the ion migration.

During the ion-current measurements
the secondary beam was slowed down in the precision SiO$_2$ degraders
and wedge and the ions were thermalized in ultra-high purity helium at
a pressure of 650~mbar. The effective thickness of the glass degraders
was varied by rotation.  The ratio of ions stopped in the chamber
to the number of incident ions was measured as function of the
degrader thickness using the silicon detector both with and without
filling gas.  The point-by-point difference between these
measurements, shown and discussed below, provided the range distributions
of the ions. After the range measurements were complete the silicon
detector was removed from the active volume.

\section{Ion Currents}\label{sec:ioncurrents}

\begin{figure} %% 5
\begin{center}
\includegraphics[width=0.95\textwidth]{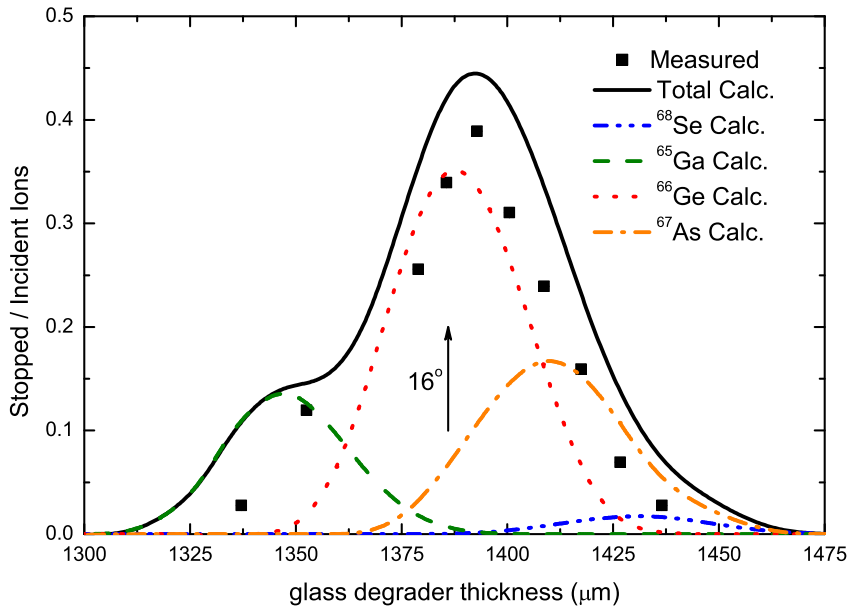}
\caption{\label{fig:range} The experimental (squares) and theoretical (lines) fraction of incident heavy ions stopped in a 39~cm long path at p=650~mbar
are shown as function of the degrader thickness. In addition, the calculated contributions from each species in the cocktail beam are indicated, see the text for details.  The glass thickness at an angle of 16$^o$ is indicated for reference.}
\end{center}
\end{figure}

Several measurements of the ion currents created by the fast beam
were obtained in the present work.  As indicated above, the stopping profile or
essentially the range distribution of all of the incident ions was
measured in the silicon detector.  The results are shown by the
squares in Fig.\ref{fig:range}.  The stopping profile was also
calculated with the TRIM code\cite{TRIM} using information on the
transverse and longitudinal momenta of the mixed secondary beam at
the degrader position predicted by a LISE++ simulation\cite{LISE}.
The predicted range distribution has no free parameters and the centroid
was brought into the agreement shown in Fig.\ref{fig:range}
by decreasing the absorber thickness by only 60~$\mu$m out of a
total thickness of $2685\pm 8\mu$m. This discrepancy is within
the uncertainty of the momentum measurement of the fragment
distribution from the A1900 separator itself.  The observed distribution
in Figure~\ref{fig:range} is
slightly narrower than the calculated distribution which may be due to an
overestimate of the acceptance of the system in the LISE++ simulation.
Note that 50~cm of helium at 650~mbar is equivalent to 25~$\mu$m of
this glass so that essentially all ions of any one of the incident isotopes can be
stopped in the gas when the degrader is centered on the maximum of
each curve in Fig.\ref{fig:range}.
%%
%  figure 5 here

\subsection{Electrons and Negative Ions}\label{sec:negativeions}
After the silicon detector was removed, a voltage distribution was
applied to electrodes inside the gas to separate the ion-electron pairs.
Electrons and any anions were collected at high potential on the
third ring (counting from the entrance window) while the cations
were drifted toward the extraction
nozzle that was near ground potential.  The drift field was
approximately constant at $\sim$4~V/cm, approximately 0.025~Td, at
a pressure of 650~mbar.   The negative ion current was
measured by an electrometer connected to the third ring electrode
that was at a slightly higher
potential than its neighbors.  The gas cell was thus behaving like
a gas-filled ionization chamber (Bragg geometry) in which the incident
ions leave a fraction of their incident energy in the gas whenever the
glass is too thin to have the ions stop in the gas.
%%
%  figure 6 here

\begin{figure}[t] %% 6
\begin{center}
\includegraphics[width=0.95\textwidth]{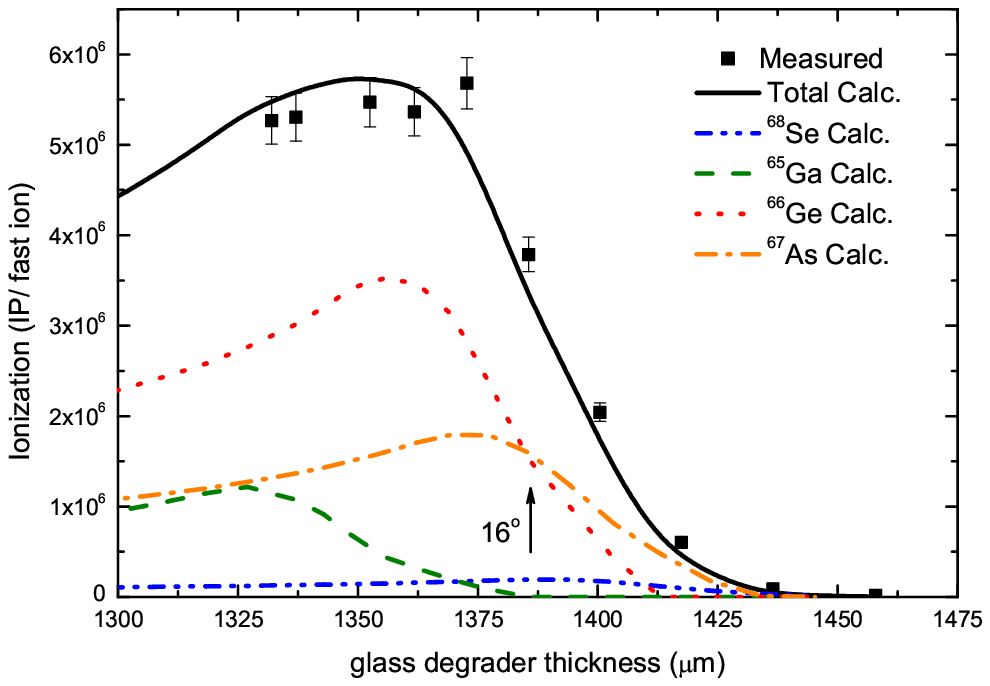}
\caption{\label{fig:ionization} The measured (squares) and theoretical (solid line) ionization profile from the negative ion current (in ion-electron pairs/incident ion) is shown as function of glass degrader thickness. The contributions from each projectile fragment is shown, similar to Figure~\protect{\ref{fig:range}}.}
\end{center}
\end{figure}

The negative ion current was obtained in DC mode as a function of
degrader thickness and divided by the incident fast-ion rate and
by the effective ionization potential of helium (42~eV) to produce
the ionization profile shown in Fig.\ref{fig:ionization}.   The
typical incident rate during this measurement was ~10$^4$~ions/s.
The ionization profile as a function of degrader thickness is a
measurement of the
Bragg curve when the negative ion collection is complete.  For
comparison, the Bragg curves were calculated as function of the
glass degrader thickness for each ion type in the gas cell with
the TRIM code using the input from the LISE++ simulations.
The experimental results and the total
ionization predicted for the mixed beam are shown in
Fig.\ref{fig:ionization} . Once again there is excellent agreement
between the total theoretical ionization and the experimental
measurement.  In the present case each incident ion creates
approximately 5x10$^6$ ion-electron pairs during the stopping process that
are distributed along a column. The stopping volume containing the
ionization for all of the fast ions was estimated to be
58~cm$^3$ from the range and diameter of the incident
beam.  Such large numbers of ion-electron pairs is typical for projectile
fragments when a substantial fraction of the range
distribution is collected, as discussed above.

\subsection{Positive Ions }\label{sec:positiveions}
The ion current due to the positive ions (cations) from the NSCL
gas cell can only be measured after the nozzle and thus outside
the gas cell.  The measurement and interpretation of the cation
current is complicated by the ion chemistry occurring during the drift
time and by any mass selection in the ion guides after the nozzle.  In
order to study the behavior of extracted ions from the gas cell,
2~ms long pulses of secondary beam were injected into the
gas cell. This pulse width is small compared to the drift time of the
cations but provides enough ionization for the measurements.  The
time profile of positive ions exiting the gas cell after the beam
pulse was recorded with a digital oscilloscope connected to a
microchannel plate (MCP) detector placed several meters downstream
from the ion guides. The MCP intercepted the ions after they
passed through the ion guides and were accelerated to $\sim$5~kV.
The next beam pulse was injected into the gas cell only after a
pause to transfer the data from the oscilloscope.

\begin{figure}[t] %% 7
\begin{center}
\includegraphics[width=0.75\textwidth]{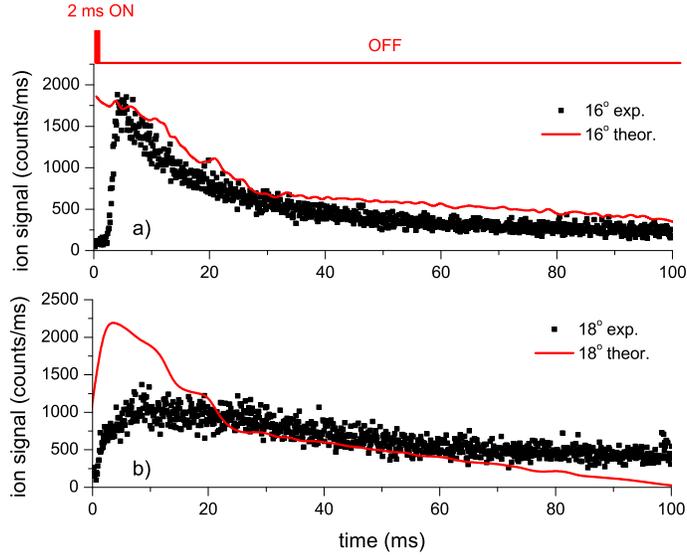}
\caption{\label{fig:T_profiles} Measured (dots) and calculated
(line) time profiles of positive ions extracted from the gas cell
after a beam pulse enters the gas cell and the glass degrader is
positioned at either (a) 16$^o$ or (b) 18$^o$.  The primary ionization
was created by 54 incident projectile fragments created in 2~ms and
was drifted with 0.01~ms time steps.}
\end{center}
\end{figure}

Examples of the measured time profiles are shown in
Fig.\ref{fig:T_profiles} (dots) for an incident beam intensity of
2.7$\cdot$10$^{4}$ pps with the glass degrader, 1.332~mm
 thick, tilted to either (a) 16$^o$ or (b)
18$^o$ with respect to the normal to the beam direction.  The
projectile fragments travel the entire length of the gas cell under
these conditions, see
Fig.\ref{fig:range}, and stop in the central spherical electrodes.
The columns of ion-electron pairs contain approximately 1.9x10$^8$ and
%%{\bf 2.7x10$^{4}$*0.002s*5x10$^{5}$ from Fig.2}
2.7x10$^7$ IP per projectile fragment (PF) distributed along the entire
central axis which would correspond to ion pair densities of
approximately
%%{\bf 1.9x10$^{8}$/58cm$^3$}
3.3x10$^6$ and
%%{\bf 2.7x10$^{7}$/58cm$^3$}
4.7x10$^5$ ion-electron pairs/cm$^3$/PF.  Recall that these densities are sufficiently low
that recombination can be neglected.  Following the discussion of charge
exchange processes above, we expect that all of the electrons will be rapidly
collected on the ring anode and that all of the positive charge will be
transferred to impurity molecules before leaving the gas cell given their
long drift times.

The ion guides were set to transmit mass-to-charge ($m/q$)
values in the range of approximately 40 to 80.  The composition of this
ion beam was not measured during the beam time but the results of many
studies of the ion output from the NSCL gas cell indicate that it
consists mainly of hydrocarbon molecular ions and small fluorinated
hydrocarbons, see Figure~\ref{fig:mass} and discussion
above. Thus, we expect the bulk of the cations extracted
from the gas cell to be transmitted to the MCP detector and recorded.
The observed time distributions peak at short times and then have
an exponential tail.  The observed time scales are consistent with
simple expectations and detailed calculations of the ion migration are
discussed in the next section.

The measured total extraction efficiency of molecular ions was obtained from
the ratio of the Integral of the measured time profiles from Fig.\ref{fig:T_profiles} of
the extracted ions to the total number of molecular ions from the primary ionization.
This efficiency turns out to be 5$\cdot 10^{-4}$ for the 16$^o$ data and 6$\cdot 10^{-4}$
for 18$^o$ data.  These ratios include losses due to the extraction efficiency, the
transport efficiency in the ion guides and the detection efficiency.  The transport
efficiency was estimated to be 0.2(1) by measuring total ion current at various
points along the ion guides and beam line.  Unfortunately the MCP was run in a
analogue mode and the constant of proportionality was not measured so that the
extraction efficiency could not be determined.
%%
%  figure 7 here

\subsection{Particle-in-cell Calculations}\label{sec:piccode}
The detailed ion migration in the gas cell will depend on the combined
effects of the applied field,  the space charge of the ions and finally by
gas flow in the region of the nozzle.
The removal of the electrons and creation of a net positive charge
in the gas cell leads to an increase of the potential in the gas
cell, which counteracts the applied field\cite{MH_NIM02} even in
the case of a short pulse of ionization\cite{CV_NIM05}. A
particle-in-cell (PIC) code was developed to model the ion motion
in this situation using cylindrical symmetry\cite{MF_HI07} without
the effects of gas flow. The few projectile fragments that created
the primary ionization were placed in the gas volume according to their
calculated range distributions, Fig.~\ref{fig:range} and the
much more numerous He$_2^+$ ions were placed according to the
calculated ionization distribution, Fig.~\ref{fig:ionization}. The
ion mobilities of all of the cations in helium are only approximately
twice that of He$^+$ (which has a very short half life, discussed above)
so that the chemical nature of each ion is not important. The migration of
all of these ions was then tracked as a function of time from
their creation until extraction or loss at the electrodes and/or
boundaries and the drift time was recorded.
%%
%  figure 8 here

\begin{figure}[b] %% 8
\begin{center}
\includegraphics[width=0.95\textwidth]{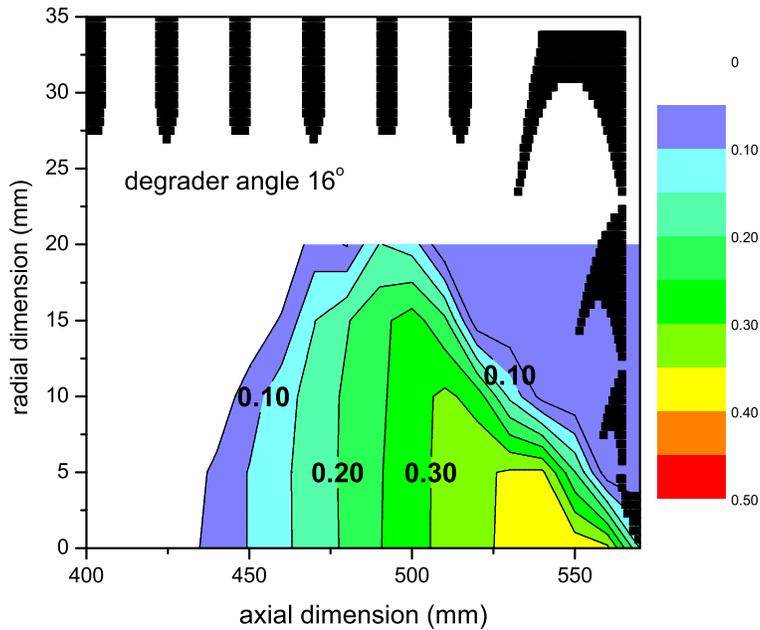}
\caption{\label{fig:extr_prob} Density plot of extraction probabilities of molecular ions from the gas cell when the rotation angle of the glass degrader was 16$^o$. The black shapes along the edge indicate the positions of the spherical electrodes and the ring electrodes numbers 16 to 21.}
\end{center}
\end{figure}

The absolute time distributions of the extracted molecular ions
after the beam pulse are shown by the solid lines in Fig.\ref{fig:T_profiles}
for different glass degrader thicknesses.
Small wiggles in the calculated curves are indications of the statistical
accuracy of the calculations. The calculated results are in good general
agreement with the observations and reflect the interplay of the shape of
the drift field and the radial expansion due to the space charge.  Recall
that the geometry of the gas
cell is such that there is a radial component to the applied field only
in the region of the spherical electrodes. The time distribution observed
with the thinner degrader (a more uniform column of ionization)
is more sharply peaked in the data than in the calculations.  This time
peak is created by the {\em loss} of ions at longer times due to the expansion
of the ion cloud in the region of the ring electrodes.  In fact, only ions created
relatively close to the extraction region are collected (see below). The more
gradual rise and fall of the number of extracted ions in the case
of lower ionization is only generally reproduced by the calculations due
to the fact that the primary ionization is higher in the region further from the
extraction region.

The extraction probability was obtained for each grid element by
following the trajectories of ions placed at each of the nodes of the PIC grid.
The extraction probability is the ratio of the number of ions that
entered the throat of the nozzle electrode to the total number of initial ions.
The resulting values were very similar for the two cases and the values for
the 16$^o$ degrader angle are shown in Fig.\ref{fig:extr_prob}.  The
extraction probability has a maximum of approximately 50\% for ions
created near the nozzle and drops to near zero for ions created more than
10~cm upstream.  Recall that this simulation does not include forces due to
gas flow that would increase the extraction efficiency near the nozzle.

The predicted extraction probability of molecular ions can be determined by
folding the extraction probabilities shown in Figure~\ref{fig:extr_prob} with
the initial ion distribution. Note that the calculations indicated that no
ions would be extracted outside the volume indicated in Figure~\ref{fig:extr_prob}.
In the present case the ionization distribution was approximately uniform
over a column of 123 cm$^3$ because the radioactive beam did not stop in the
gas.  Combining these distributions one obtains a predicted extraction
probability of 8\% that is dominated by the region near the nozzle.  This
value of a few percent is similar to the values previously reported for the
extraction of radioactive ions injected into the NSCL gas cell at low
rates\cite{LW_NIM05}.

\section{Summary and Conclusions}\label{sec:sum}
The fates of various ions created by the injection of pulses
of projectile fragments into the NSCL gas cell were considered.
A discussion of the gas phase chemistry indicated the importance
of charge exchange reactions for conversion of the primary ions (He$_2^+$)
into various molecular ions. The lifetimes for these charge exchange
reactions are such that all of the primary ionization will be transferred to molecules
before extraction in this system. This conversion essentially
increases the mass of the moving ions from A=8 into the mass region
of the injected projectile fragments and obviates mass-selection
against the primary ionization in this case.

The measured distributions of the incident ions and the primary
ionization were shown to be in excellent agreement with the
predictions. A Particle-In-Cell code was developed to describe
the motion of the positive ions in the gas cell including a
self-consistent space charge. The results of the calculations are in
good general agreement with the experimental data.

The assistance of the LEBIT group during the acquisition of the
time spectra is gratefully acknowledged.  This work was supported
in part by the US Department of Energy under grant
DE-FG02-00ER41144 and by the US National Science Foundation Grant
PHY-06-06007.

% NIM wants figures at the end of document \section{Figures}

\end{document}